\documentclass[12pt]{article}

\usepackage{graphicx}
\usepackage{fancybox}
\usepackage{amsmath}
\usepackage{amssymb}
\usepackage{latexsym}
\usepackage{epsfig}

\newcommand{\be}{\begin{equation}}
\newcommand{\ee}{\end{equation}}
\newcommand{\ea}{\end{eqnarray}}
\newcommand{\ba}{\begin{eqnarray}}
\newcommand{\no}{\nonumber}

\newcommand{\Th}[2]{\Theta_{#1,#2}}
\renewcommand{\th}{{\theta}}

\newcommand{\nn}{\nonumber\\}

\newcommand{\br}{\Bbb R}
\newcommand{\bz}{\Bbb Z}

\newcommand{\hcK}{\widehat{\cal K}}

\newcommand{\msc}[1]{\mbox{\scriptsize #1}}

\newcommand{\bsz}{\msc{{\bf Z}}}

\newcommand{\cN}{{\cal N}}

\newcommand {\eqn}[1]{(\ref{#1})}

\makeatletter
\@addtoreset{equation}{section}
\def\theequation{\thesection.\arabic{equation}}
\makeatother

\def\Z{\Bbb Z} \def\t{\theta} \def\z{\zeta} \def\a{\alpha}
\def\={\;=\;} \def\O{\text O}

\newcommand{\bh}{{\Bbb H}}

\begin{document}

\begin{titlepage}
 \
 \renewcommand{\thefootnote}{\fnsymbol{footnote}}
 \font\csc=cmcsc10 scaled\magstep1
 {\baselineskip=14pt
 \rightline{
 \vbox{\hbox{YITP-08-12}
       \hbox{UT-08-04}
       }}}

 \baselineskip=20pt
\vskip 1cm
 
\begin{center}

{\bf \Large  Modular Forms and Elliptic Genera for ALE Spaces} \footnote
{Contribution to the Proceedings of the workshop in honour of Professor Tsuchiya's retirement, Nagoya  University, March 2007}

 \vskip 1.5cm
\noindent{ \large $ \mbox{Tohru Eguchi}^{\,\dag,\,\ddag}$},
{\large $\mbox{Yuji Sugawara}^{\,\ddag}$} \\

and 

{ \large $\mbox{Anne Taormina}^{\, \sharp}$} \\

\vskip 1cm


{\it Yukawa Institute for Theoretical Physics, \\
Kyoto University, Kyoto, Japan 606-8502}$^{\,\dag}$,

\vskip 5mm
 
{\it Department of Physics,  Faculty of Science, \\
  University of Tokyo, Tokyo, Japan  113-0033}$^{\,\ddag}$,

\vskip 5mm

{\it Department of Mathematical Sciences, University of Durham, \\
South Road, DH1 3LE Durham, England}$^{\,\sharp}$

\end{center}

\bigskip

\begin{abstract}
When we describe string propagation on non-compact or singular Calabi-Yau manifolds by
CFT, continuous as well as discrete representations 
appear in the theory. 
These representations mix in an intricate way under the modular
 transformations. In this article, we propose a method 
of combining  discrete and continuous representations so that 
the resulting combinations have a simpler modular behavior
 and can be used as conformal blocks of the theory.
We compute elliptic genera of ALE spaces and obtain
results which agree with those suggested from 
the decompactification of K3 surface. Consistency 
of our approach is assured by some remarkable 
identity of theta functions. 

We include in the appendix some new materials on the representation theory  of 
${\cal N}=4$ superconformal algebra.
\end{abstract}

\vfill

\setcounter{footnote}{0}
\renewcommand{\thefootnote}{\arabic{footnote}}

\end{titlepage}

\baselineskip 18pt

\newpage

\section{Introduction}

Description of strings propagating on non-compact curved background is a challenging 
problem in particular when the space-time develops a singularity. A better grasp of underlying 
conformal field theory (CFT) should shed light on the physics of such space-time.

When a Calabi-Yau (CY) manifold is non-compact or singular, it is
necessary to introduce a CFT possessing continuous as well as
discrete representations in order to describe its geometry. These
CFT's have a central charge above the "threshold", i.e. $c=3$ for
${\cal N}=2$ supersymmetric case, and are of non-minimal type. We
may call these theories generically as Liouville type theories. Since
continuous and discrete representations mix under modular
transformations, representations of Liouville theories in general do not have
good modular properties. Thus it is a non-trivial problem to
construct suitable modular invariants describing the geometry of
non-compact CY.

In this paper we present an attempt at constructing (holomorphic)
modular invariants for some non-compact CY manifolds. In particular
we propose the elliptic genera for the ALE spaces which are the degenerate limits of K3 surface. 
It turns out that the consistency of our approach hinges on the validity
of some theta-function identities. These non-trivial identities have
been proved by D.Zagier and the proof is given in section 2.5. 

This paper is a contribution to the Proceedings of the workshop in honor of prof. A. Tsuchiya's retirement from  Nagoya University on March 2007. It is an expanded version of ref \cite{EST} and contains some new materials on the representation theory of superconformal algebras, in particular on higher level ${\cal N}=4$ character formulas. 

\subsection{Bosonic Liouville theory}

We start our discussions by reviewing the simple case of bosonic
Liouville theory. Its stress tensor is given by \ba T(z)= -{1\over
2}(\partial\phi)^2+{{\cal Q}\over 2}\partial^2\phi \ea where ${\cal
Q}$ is the background charge. Central charge is given by \ba
c=1+3{\cal Q}^2. \ea If we parameterize ${\cal Q}$ as ${\cal
Q}=\sqrt{2}(b+{1/b})$, the vertex operator \ba \exp(\sqrt{2}b\phi)
\ea has a conformal dimension $h=1$. Liouville theory is defined as
a theory perturbed by this marginal operator (Liouville potential)
from free fields.

Dynamics of boundary Liouville theory became clarified in late
1990's by the method of conformal bootstrap \cite{FZZTZZ}. We first
reintroduce the result of conformal bootstrap using representation
theory and the modular properties of character formulas.

It is known that there are two types of representations in bosonic
Liouville theory: continuous and identity representation. Their
character formulas and their S-transformation are given 
by $(q\equiv e^{2\pi i \tau})$\\

continuous representations;\hskip1cm $p>0$ \ba &&
\chi_p(\tau)={q^{h-{c\over 24}}\over \prod_{n=1}(1-q^n)}={q^{{p^2
\over 2}}\over \eta(\tau)}
,\hskip1cm h={p^2\over 2}+{{\cal Q}^2\over 8}\no\\
&&\no\\
&&\chi_p(-{1\over \tau})=2\int_0^{\infty}dp' \cos(2\pi
pp')\chi_{p'}(\tau), \label{cont}\ea

\bigskip
 identity
representation;\hskip2cm $h=0$ \ba
&&\chi_{h=0}(\tau)={q^{-{{\cal Q}^2\over 8}}(1-q)\over \eta(\tau)},\no\\
&&\no \\
&&\chi_{h=0}(-{1\over \tau})=4 \int_0^{\infty} dp \sinh(2\pi
bp)\sinh({2\pi p\over b})\chi_p(\tau)\label{disc} \ea

We identify the LHS of the above equations as describing the open
string channel and RHS as the closed string channel. We then find
that open and closed channels have different spectra:\\

$\begin{array}{ll}
\mbox{open} & \hskip3cm \mbox{closed} \\
\null \\
\left\{\begin{array}{l}\mbox{continuous rep.} \\
\null \\ \mbox{identity rep.}\end{array}\right.
& \hskip3cm \mbox{continuous rep.} \\
\end{array}$
\\

\vspace*{2mm} \noindent Namely, there exist no identity
representation in the closed string channel. This is consistent with
the presence of mass gap and the decoupling of gravity in
non-compact space-time. Indeed the conformal dimension of a vertex operator
$e^{\alpha \phi}$ is given by \ba
h(e^{\alpha\phi})&=&-{\alpha^2\over 2}+{\alpha {\cal Q}\over 2}=
-{(\alpha  -{1\over 2}Q)^2 \over 2}+ {{\cal Q}^2\over 8}\no \\
&=&{p^2\over 2}+{{\cal Q}^2\over 8}\ge {{\cal Q}^2\over 8} \hskip3mm
\mbox{ for} \hskip3mm \alpha=ip+{1\over 2}{\cal Q}\label{gap} \ea
for continuous representations. Thus there is a gap of ${\cal
Q}^2/8$ in the spectrum of continuous representations.

Let us next turn to the brane-interpretation of transformations
(\ref{cont}), (\ref{disc}). We introduce ZZ and FZZT brane boundary
states $|ZZ\rangle,\,|FZZT\rangle$ and identify the character
functions as the inner product \ba
&&\chi_0(-{1\over \tau})=\langle ZZ|e^{i\pi \tau H^{(c)}}|ZZ\rangle\label{zz} \\
&&\chi_p(-{1\over \tau})=\langle FZZT;p\,|e^{i\pi \tau
H^{(c)}}|ZZ\rangle\label{fzzt}
\ea\\
where $H^{(c)}=L_0+\bar{L}_0- \frac{c}{12}$ 
is the closed string Hamiltonian.
Using Ishibashi states $|p\,\rangle\rangle$ with momentum $p$ which
diagonalize the closed string Hamiltonian \ba \langle\langle
p\,|e^{i\pi \tau
H^{(c)}}|p'\rangle\rangle=\delta(p-p')\chi_p(\tau)\label{ishibashi}
\ea boundary states are expanded as \ba
&&|ZZ\rangle=\int_0^{\infty} dp\,\Psi_0(p)\,|p\,\rangle\rangle \label{zz-state} \\
&& \no \\
&&|FZZT;p \,\rangle=\int _0^{\infty}{dp'}
\,\Psi_p(p')\,|p'\rangle\rangle. \label{fzzt-state} \ea 
We then have
\ba
&& |\Psi_0(p)|^2=4 \sinh \sqrt{2}\pi pb 
\sinh{\sqrt{2}\pi p\over b}\,, \\
&& \null \no \\ &&\Psi_p(p')^*\Psi_0(p')=2 \cos 2\pi pp'.  
\ea 
Solving these relations one finds the boundary wave-functions 
\ba &&
\Psi_0(p)={2\sqrt{2}\pi ip \over \Gamma(1+i\sqrt{2}pb) 
\Gamma(1+{i\sqrt{2}p\over b})},\\
&&\Psi_p(p')={-1\over \sqrt{2}\pi ip'}
\Gamma(1-\sqrt{2}ibp') \Gamma(1-{\sqrt{2}ip'\over
b})\cos(2\pi pp'). \ea Up to phase factors the above results agree
with those of conformal bootstrap \cite{FZZTZZ}.\\

\bigskip

\subsection{${\cal N}=2$ Liouville theory}

For the sake of applications to string theory let us now consider
${\cal N}=2$ supersymmetric version of Liouville theory. In ${\cal
N}=2$ system possesses two bosons, one of them coupled to background
charge and the other one is a compact boson, and two free
fermions. It is known that ${\cal N}=2$ Liouville theory is T-dual
to $SL(2;\br)/U(1)$ supercoset model which describes the
space-time of the two-dimensional black hole \cite{HK}. In general ${\cal
N}=2$ Liouville is geometrically interpreted as describing the
radial
direction of a complex cone.\\

In the following we concentrate on the case when ${\cal N}=2$
Liouville has a central charge \ba &&\hat{c}={c\over 3}=1+{2 \over
N},  \hskip2mm  {\cal Q}=\sqrt{2/N}.\no \ea
for the sake of simplicity.  We denote this case as the model  $L_N$. 
Here $N$ is an arbitrary
positive integer. This theory is T-dual to two-dimensional black
hole with an asymptotic radius of the cigar $\sqrt{2N}$.

Unitary representations of ${\cal N}=2$ superconformal algebra with
$\hat{c}=1+{2\over N}$
are given by\\

$\left\{\begin{array}{llll}
\mbox{identity rep.} & h=0, & j=0 & \hskip1cm \mbox{vacuum} \\
&\\
\mbox{continuous reps.} & p>0, & j={1\over 2}+i{p\over {\cal Q}}
& \hskip1cm \mbox{non-BPS states}\\
& & &\\
\mbox{discrete reps.} & 1\le s \le N, & j={s\over 2}& \hskip1cm
\mbox{BPS states, chiral primaries}
\end{array}\right.$\\
\\
\vspace*{1mm} Here $p$ and $s$ label continuous and discrete
representations of ${\cal N}=2$ Liouville theory, respectively.
${\cal N}=2$ representations are in one to one correspondence with
those of level $k=N$
$SL(2;\br)/U(1)$ coset theory with the value of spin $j$ indicated as above.  \\

In applications to string theory we consider the sum over
spectral flows of each ${\cal N}=2$ representation and define an
extended character \cite{ES1}\footnote
   {Here the spectral flow is summed over modulo $N$ for the sake of convenience.
Idea of extended character has been introduced in \cite{ET} where the    
irreducible characters of $\cN=4$ algebra are identified as extended characters of $\cN=2$ algebra. For related works  see \cite{Odake,Miki,IPT}. } 
\ba \chi^{NS}_*(r;\tau,z)=\sum_{n\in\,
r+N \bz}q^{{\hat{c}\over 2}n^2}e^{2\pi i\hat{c}zn}
ch^{NS}_*(\tau;z+n\tau)
\label{irred-extended}
\ea Here $ch^{NS}_*(\tau;z)$ denotes an irreducible character of ${\cal
N}=2$ superconformal algebra (in NS sector). 
Extended characters carry some
additional label
 \begin{eqnarray}
 &&1. \mbox{ Identity representations}:  \nonumber \\
&&\chi^{NS}_{id}(r;\tau,z); \hskip1cm  r\in \bz_N, \\
&& \nonumber \\
&& 2.\mbox{ Continuous representations}: \\
&&\chi^{NS}_{cont}(p,\alpha;\tau,z);  \no \\
&& \nonumber \\
&&3. \mbox{ Discrete representations}:  \\
&& \chi^{NS}_{dis}(s,s+2r;\tau,z); \, \hskip1cm  r \in \bz_N, 1\le s\le
N\nonumber
\end{eqnarray}\\
Explicit form of these characters 
are presented in the Appendix A. 
We also present the form of modular transformation.
Here we recall that the S
transform of these functions has the following pattern \ba
&&\mbox{(continuous rep)}\stackrel{S}{\longrightarrow}
\mbox{(continuous rep)}\label{cont-s} \\
&& \no  \\
&&\mbox{(identity rep)}\stackrel{S}{\longrightarrow} \mbox{(discrete
rep)}+ \mbox{(continuous rep)} \label{iden-s} \\
&&  \no \\
 &&\mbox{(discrete
rep)}\stackrel{S}{\longrightarrow} \mbox{(discrete rep)}+
\mbox{(continuous rep)} \label{disc-s} \ea Namely, a continuous representation transforms
into an integral over continuous representations while an identity and
discrete representation transforms into a sum of discrete representation and an
integral over continuous representations.
Such a pattern was first observed in ${\cal N}=4$ representation theory \cite{ET}.\\

\begin{enumerate}
\item
As in the bosonic Liouville theory, there appear no identity 
representations in the RHS of above formulas.
\item
While the identity representation disappears 
after a first S-transform, it comes back 
after a 2nd transform: 
this happens when one deforms the contour of momentum integration  for the sake of convergence 
and picks up a pole 
in the complex plane corresponding to the identity representation.
It is further possible to check that  
$S^2=C$ and $(ST)^3=C$, where
$C$ is a charge conjugation matrix
which acts as $C\,:\, (\tau,z)\, \rightarrow \, (\tau,-z)$.
\end{enumerate}

As compared with the case of minimal theories where only discrete representations exist which rotate into each other under the S-transform, the above transformation laws (\ref{cont-s})-(\ref{disc-s}) are much more complex and in particular discrete  representations mix with continuous representations. We can check that even under the transformation $ST^2S^{-1}$ one can not eliminate the contribution of continuous representations in the transform of discrete representations ($ST^2S^{-1}$ is a generator of $\Gamma(2)$ which is the subgroup of 
$SL(2;\bz)$ keeping the spin-structure fixed). 
It seems not possible to
eliminate the mixing of continuous representations 
under any subgroup of the modular group.

\vskip1cm

We have three types of boundary states of ${\cal N}=2$ theory
corresponding to each representation. The boundary wave functions
are again given by the elements of the modular $S$ matrix. We can
compare our expressions with known results of $SL(2;\br)/U(1)$ theory
obtained by semi-classical method using the geometry of 2d black
hole and DBI action. It is found \cite{ES1,Packman} 
that ${\cal N}=2$ theory reproduces essentially 
the correct wave functions of D-branes 
of 2d black hole \cite{RS}. Thus the representation theory seems 
quite consistent with the semi-classical analysis.
However, the character formulas themselves do not have good modular
properties and it is a non-trivial problem to construct conformal
blocks with good modular behaviors.\\

\section{Geometry of ${\cal N}=2$ Liouville Fields}

Let us now consider models of the following type: tensor product 
of ${\cal N}=2$ Liouville theory $L_N$ (of $\hat{c}=1+\frac{2}{N}$)
and ${\cal N}=2$ minimal model $M_k$ with level $k$
\cite{OV}
\ba
&& L_{N}\otimes M_{k}. 
\ea
If we choose \ba N=k+2 \ea the central
charge becomes integral 
\ba c_L+c_{{M}}=3(1+{2\over N})+3(1-{2\over
k+2})=6 \ea 
and the theory (after $Z_N$ orbifolding) describes (complex) 2 dimensional CY
manifolds. They are identified as the (A-type) ALE spaces which are obtained by blowing up $A_{N-1}$ singularities \cite{OV}. 
At $N=1$ (without minimal model), we have $\hat{c}=3$ and
the space-time of a conifold \cite{GV}.
We may as well consider the tensor products of Liouville theories
and minimal models. These describe other singular geometries
like ${A_{N-1}}$ spaces fibered on $P^1$ etc. 
\cite{Lerche,HK2,ES3}\\

\subsection{Elliptic genus and CY/LG correspondence}

The elliptic genus is defined by taking 
the sum over all states in the
left-moving sector of the theory while the right-moving sector is
fixed at the Ramond ground states;
\ba &&Z(\tau,z)=Tr_{R\otimes
R}(-1)^{F_L+F_R} 
e^{2\pi i z J_o^L} 
q^{L_0-\frac{\hat{c}}{8}}
\bar{q}^{\bar{L}_0-\frac{\hat{c}}{8}}.  
\label{elliptic genus def}
\ea
Here $J_0^L$ denotes the $U(1)_R$ 
charge in the left-moving sector.
The trace is taken in the Ramond-Ramond sector. 
At specific values of
$z$ we have \ba
&&Z(\tau,z=0)=\chi,\hskip4.5cm \mbox{Euler number}\no \\
&&Z(\tau,z={1/2})=\sigma+{\cal O}(q), \hskip2.7cm \mbox{Hirzebruch signature} \no\\
&&Z(\tau,z={(\tau+1)/2})=\hat{A}q^{-1/4}+{\cal O}(q^{1/4})
,\hskip0.3cm \mbox{$\hat{A}$ genus}\no \ea
The elliptic genus is an
invariant under smooth variations of the parameters of the theory and is
useful, for instance, in counting the number of BPS states. We
compute the elliptic genus of a non-compact CY manifolds by pairing
the Liouville theory with ${\cal N}=2$ minimal
models.\\

Before going into the computation of elliptic genera we first recall
the results of CY/LG correspondence 
\cite{Witten-CY/LG}. 
We consider a Landau-Ginzburg
(LG) theory with a superpotential \ba W=g\,(X^{k+2}+Y^2+Z^2)  \ea
which in the infra-red limit acquires scale invariance and
reproduces the ${\cal N}=2$ minimal theory with $\hat{c}=1-{2/k}$.

 In the ${\cal N}=2$ minimal theory $M_{N-2}$,
the contribution to elliptic genus comes from
 the Ramond ground states
\ba
&&Z_{\msc{minimal}}(\tau,z)=\sum_{\ell=0}^{N-2}ch^{\tilde{R}}_{\ell,\ell
+1}(\tau;z). \label{minimal} \ea 
Here $ch^{\tilde{R}}_{\ell,\ell+1}(\tau;z)$ denotes the character
of minimal model $M_k$ associated to 
the Ramond ground state labeled by $\ell=0,1,\ldots, N-2$.
See e.g. \cite{Gepner,RY} for their explicit expressions.
$\tilde{R}$ denotes 
the Ramond sector with $(-1)^F$ insertion.
On the other hand as the coupling
parameter is turned off $g\rightarrow 0$, LG theory becomes a
free theory of chiral field $X$ with $U(1)_R$ charge $={1/N}$. Thus
the theory possesses a free boson of charge ${1/N}$ and free fermion
of charge ${1/N-1}$. Combining these contributions one
obtains \cite{Witten} \ba Z_{LG}(\tau,z)={\theta_1(\tau,(1-{1\over
N})z)\over \theta_1(\tau,{1\over N}z)}. \label{LG} \ea These two expressions 
(\ref{minimal}),(\ref{LG}) in fact agree 
with each other \ba
Z_{\msc{minimal}}=Z_{LG}. \ea

We would like to try a similar construction in Liouville sector
as in the case of minimal models. 
Ramond ground states corresponds to the extended 
discrete characters;
\ba \chi^{\tilde{R}}_{dis}(s,s-1;\tau,z),
\hskip1cm s=1,\cdots,N \ea 
and the elliptic genus is expressed as 
their sum, which is explicitly evaluated in \cite{ES2}
as follows;
\footnote
  {Precisely speaking, in \cite{ES2} we adopt a slightly different convention for 
   the `boundary contribution $(s=1,N+1)$' of discrete representations, which 
   yields the anti-symmetrized Appell function (\ref{anti-symm})
$$
Z_{\msc{Liouville}} = -\widehat{{\cal K}}_{2N} (\tau,z) 
\frac{i\theta_1(\tau,z)}{\eta(\tau)^3}
$$
rather than \eqn{Z L}. However, the difference drops off 
in the orbifold procedure \eqn{orbifold} \cite{ES2}. 
Namely, one may replace ${\cal K}_{2N}(\tau,z)$ with 
$\widehat{{\cal K}}_{2N}(\tau,z)$ in \eqn{orbifold}.
 }
 \ba
Z_{\msc{Liouville}}&=& - \sum_{s=1}^{N}
\chi_{dis}^{\tilde{R}}(s,s-1;\tau,z)\no \\
&=& - {\cal K}_{2N}(\tau,{z\over N}){i\theta_1(\tau,z)\over
\eta(\tau)^3} 
\label{Z L}
\ea 
Here we have introduced the notation of an
Appell function ${\cal K}_{k}$ \cite{STT,Pol}\\
\ba {\cal K}_{k}\,(\tau,z)\equiv \sum_{n\in \bz}{q^{{ k\over 2}n^2}
y^{kn}\over 1-yq^n}\,,\hskip1cm y=e^{2\pi i z}.\label{appell} \ea
We also use the anti-symmetrized version of Appell function 
defined as
\ba
\widehat{{\cal K}}_k(\tau,z)&
\equiv &{1\over 2}\left({\cal K}_k(\tau,z)-{\cal K}_k(\tau,-z)\right)
\equiv {\cal K}_k (\tau,z) - \frac{1}{2} 
\Theta_{0,\frac{k}{2}} (\tau,2z)\label{anti-symm}
\ea 
Unlike the theta
functions of the minimal models, 
the Appell function in Liouville theory
does not have a good modular transformation law \cite{STT}.
Complication comes from the non-trivial
denominator of the function (\ref{appell}) which arises due to
existence of fermionic singular vectors in BPS (short)
representations.


The Appell function is closely related to the function used by Miki in
\cite{Miki}: they are transformed to each other by spectral flow.
The Appell function corresponds to an expression in $\tilde{R}$ sector while Miki's function is 
in NS sector.


When we couple minimal and Liouville theory to compute elliptic
genera of $A_{N-1}$ spaces, we may use the orbifoldization 
procedure \cite{KYY} and we find \cite{ES2} 
\ba
Z_{ALE(A_{N-1})}(\tau,z)\no 
&=&{1\over N}\sum_{a,b\in{\bz_N}}q^{a^2}e^{4\pi i
az}Z_{\msc{minimal}}(\tau,z+a\tau+b)
Z_{\msc{Liouville}}(\tau,z+a\tau+b)\no \\
&&\null \no \\
&=&- {1\over N}\sum_{a,b\in \bz_N}q^{{a^2\over 2}}e^{2\pi i
az}(-1)^{a+b}{\theta_1(\tau,{N-1\over N}(z+a\tau+b))
\over \theta_1(\tau,{1\over N}(z+a\tau+b))}\no\\
&&\hskip2cm \times {\cal K}_{2N}(\tau,{1\over N}(z+a\tau
+b)){i\theta_1(\tau,z)\over \eta(\tau)^3}\label{orbifold} \ea 
In the special case of $N=2$ we have ($y\equiv e^{2\pi iz}$) \ba
&&Z_{ALE(A_1)}(\tau,z)=- \sum_{n\in \bz} (-1)^n{q^{{1\over
2}n(n+1)}y^{n+{1\over 2}}\over 1-yq^n}{i\theta_1(\tau,z)\over
\eta(\tau)^3}
\left(\,\equiv  ch_0^{\tilde{R}}(I=0;\tau,z)\,\right) .
\label{Z ALE A1}
\ea
This formula coincides with a massless character 
of ${\cal N}=4$ algebra \cite{ET}.
Unfortunately these formulas do not have well-behaved modular
properties 
and we must make a suitable modification. 

The elliptic genus
is associated with a conformal field theory defined on the torus and
hence it must be invariant under $SL(2;\bz)$ or under one of its
subgroups. Since we are dealing with superconformal field theory, it
seems natural to demand invariance under the subgroup $\Gamma(2)$
which leave fixed the spin structures \ba
\Gamma(2)=\left\{\left(\begin{array}{cc}
                                 a & b\\
                                 c & d \end{array}\right)
                                 \in SL(2;\bz),\, a=d=1,b=c=0
                \mbox{\hskip3mm mod{\hskip2mm 2}}\right\}\nonumber
\ea\\
It is known that $\Gamma(2)$ is generated by $T^2$ and $ST^2S^{-1}$.
In the following we construct elliptic genera which are invariant
under $\Gamma(2)$.

\subsection{Elliptic genus of K3}

A hint for our construction comes from the study of elliptic genus
of K3 surface (we denote 
$\theta_i(\tau) \equiv \theta_i(\tau,0)$)
\ba 
Z_{K3}(\tau,z)&=&8\left[\left(\theta_3(\tau,z)\over
\theta_3(\tau)\right)^2+\left(\theta_4(\tau,z)\over
\theta_4(\tau)\right)^2+\left(\theta_2(\tau,z)\over
\theta_2(\tau)\right)^2\right]. 
\label{K3elliptic}
\ea This formula
can be easily derived by orbifold calculation on $T^4/\bz_2$
\cite{EOTY} or by using  LG theory and LG/CY correspondence. 
One can check $Z_{K3}(z=0)=24, Z_{K3}(z=1/2)=16+...,
Z_{K3}(z=(\tau+1)/2)=-2q^{-1/4}+...$ and $Z_{K3}$ 
reproduces classical
topological invariants, $\chi$=24,\,$\sigma$=16 and $\hat{A}=-2$.

In the case of K3 surface the manifold has a hyperK\"aler structure
and the CFT possesses an ${\cal N}=4$ symmetry. Thus one can use the
representation theory of ${\cal N}=4$ superconformal 
algebra \cite{ET}.

At $\hat{c}=2$ ${\cal N}=4$ theory contains an $SU(2)$ current
algebra at level $1$. Unitary representations of ${\cal N}=4$
algebra in the NS sector are given by 
\ba &&\mbox{massive rep.}:
ch^{NS}(h,I=0;\tau,z)=q^{h-{1\over 8}}\,
{\theta_3(\tau,z)^2\over \eta(\tau)^3}, \\
&&\mbox{massless rep.}: ch_0^{NS}(I=0;\tau,z),\hskip3mm
ch_0^{NS}(I=1/2;\tau,z). \ea
 Massive representations exist only for isospin
$I=0$ and are analogous to continuous representations of ${\cal N}=2$. 
The $I=0$
and $I=1/2$ massless representations are analogues of identity and
discrete representations. There exists a relation among them \ba
&&ch_0^{NS}(I=0)+2ch_0^{NS}(I=1/2)= ch^{NS}(h=0,I=0)
\label{relation} \ea which shows that the (non-BPS) massive
representation becomes reducible as $h\rightarrow 0$ and splits into
a sum of massless (BPS) representations.

There are various ways of writing the massless characters , however,
particularly convenient expressions for 
our discussion are given by \cite{EOTY}
\ba 
ch_0^{NS}(I=1/2,\tau,z)&=&-\left({\theta_1(\tau,z)\over
\theta_3(\tau)}\right)^2+h_3(\tau)\left({\theta_3(\tau,z)\over
\eta(\tau)}\right)^2,\label{massless1}\\
&=&\left({\theta_2(\tau,z)\over
\theta_4(\tau)}\right)^2+h_4(\tau)
\left({\theta_3(\tau,z)\over
\eta(\tau)}\right)^2,\label{massless2}\\
&=&-\left({\theta_4(\tau,z)\over
\theta_2(\tau)}\right)^2+h_2(\tau)\left({\theta_3(\tau,z)\over
\eta(\tau)}\right)^2 , 
\label{massless3}\ea 
where the functions
$h_i(\tau),\,i=2,3,4$ are defined by \ba
&&h_3(\tau)={1\over\eta(\tau)\theta_3(\tau)}\sum_{m\in \bz}
{q^{m^2/2-1/8}\over 1+q^{m-1/2}}\,,\\
&&h_4(\tau)={1\over \eta(\tau)\theta_4(\tau)}\sum_{m \in \bz}
{q^{m^2/2-1/8}(-1)^m\over 1-q^{m-1/2}}\,,\\
 &&h_2(\tau)={1\over
\eta(\tau)\theta_2(\tau)}\sum_{m \in \bz} 
{q^{m^2/2+m/2}\over 1+q^m}\,.
 \ea
We note that $h_i$'s obey identities \cite{Wendland} 
\ba \hskip-1cm
h_3(\tau)-h_4(\tau)={1\over 4}\left({\theta_2(\tau)\over
\eta(\tau)}\right)^4\hskip-2mm,\hskip5mm h_2(\tau)-h_3(\tau)={1\over
4}\left({\theta_4(\tau)\over \eta(\tau)}\right)^4\hskip-2mm
,\hskip5mm h_2(\tau)-h_4(\tau)={1\over 4}\left({\theta_3(\tau)\over
\eta(\tau)}\right)^4\hskip-2mm .
\ea

Now using (\ref{massless1}-\ref{massless3}) 
we can rewrite K3 elliptic genus as\\
\ba \hskip-5mm q^{{1\over  4}}y^{-1}Z_{K3}(\tau,z')
&=&8\left[-\left(\theta_1(\tau,z)\over
\theta_3(\tau)\right)^2+\left(\theta_2(\tau,z)\over
\theta_4(\tau)\right)^2-\left(\theta_4(\tau,z)\over
\theta_2(\tau)\right)^2\right] \nonumber\\
&=& 24ch_0^{NS}(I=1/2;z)-8\hskip-2mm \sum_{i=2,3,4}
h_i(\tau){\theta_3(\tau,z)^2\over
\eta(\tau)^2}, \\
&&
\hskip4cm (\, z'\equiv z-(\tau/2+1/2)\,) 
\no \ea If one considers the
product of $\eta(\tau)$ times the sum of $h_i(\tau)$ functions \ba
8\,\eta(\tau)\hskip-2mm\sum_{i=2,3,4}
h_i(\tau)=q^{-1/8}\left[2-\sum_{n=1}^{\infty} a_nq^n\right] 
\label{q-1/8} \ea
one finds that the coefficients $a_n$ of $q$-expansion are positive
integers. Then using the relation (\ref{relation}) we can rewrite
$Z_{K3}$ into a sum of irreducible characters 
\ba && \hspace{-1.5cm}
q^{{1\over 4}}y^{-1}Z_{K3}(\tau,z')=20ch_0^{NS}(I=1/2;\tau,z)
-2ch_0^{NS}(I=0;\tau,z)+\sum_{n=1}^{\infty}a_n
ch^{NS}(h=n;\tau,z).
\ea 

Under the spectral flow from NS to R sector the $I=0$ and 1/2
representations turn into the $I=1/2$ and 0 representations, respectively. Thus the coefficient 
$-2$ in front of $ch_0^{NS}(I=0)$ in the above formula comes from the
multiplicity of the ground states of Ramond $I=1/2$ representation in the
right-moving sector. Therefore the net multiplicity of $I=0$
massless representation is 1. Hence in the NS sector the theory contains
 \ba
&&
1 \hskip0.8cm I=0 \hskip5mm \mbox{rep.}\no \\
&& 20 \hskip8mm I=1/2 \hskip5mm \mbox{reps.}\no \\
&& \infty \hskip5mm \mbox{      of massive
reps.}\,\,(h=1,2,\cdots)\no \ea
 $I=0$ NS representation corresponds to the
gravity multiplet and $I=1/2$ NS representation corresponds to matter
multiplets (vector in IIA, tensor in IIB). This is the well-known
field content in the supergravity description of string theory
compactified on K3 \cite{Seiberg}. Note that the values of the
dimension $h$ of massive representations are quantized at positive integers.
This is consistent with the T-invariance of the elliptic genus.

Now let us throw away the gravity multiplet so that we can
decompactify K3 into a sum of ALE spaces; it is known that K3 may be
decomposed into a sum of 16 $A_1$ spaces \cite{Page}.
Decompactification corresponds to dropping $I=0$ massless
representation. $I=0$ representation comes from $q^{-1/8}$ piece in
(\ref{q-1/8}) which in turn originates from the
$(\theta_2(\tau,z)/\theta_2(\tau))^2$ term in (\ref{K3elliptic}). This suggests the elliptic genus of the decompactified $K_3$ 
\ba Z_{K3, \msc{decompactified}}=8\left[\left(\theta_3(\tau,z)\over
\theta_3(\tau)\right)^2+\left(\theta_4(\tau,z)\over
\theta_4(\tau)\right)^2\right]. \ea\\

\subsection{Elliptic genera of ALE spaces}

We now propose the following formula for the elliptic genus of the
$A_1$ space \ba Z_{{A_{1}}}(\tau,z)={1\over
2}\left[\left(\theta_3(\tau,z)\over
\theta_3(\tau)\right)^2+\left(\theta_4(\tau,z)\over
\theta_4(\tau)\right)^2\right]. 
\label{A1 genera}\ea Note that in the NS sector we
have the decomposition 
\begin{eqnarray}
q^{\frac{1}{4}}y^{-1}Z_{A_1}(\tau, z')
&= &  
\frac{1}{2}\left[\left(\frac{\theta_2(\tau,z)}{\theta_4(\tau)}\right)^2
-\left(\frac{\theta_1(\tau,z)}{\theta_3(\tau)}\right)^2
\right]  \nonumber \\
&=&
ch_0^{NS}(I=1/2;\tau, z)
-{1\over
2}\eta(\tau)\left(h_3(\tau)+h_4(\tau)\right){\theta_3(\tau,z)^2\over
\eta(\tau)^3} \nonumber \\
&\equiv & 
ch_0^{NS}(I=1/2;\tau,z)+\sum_{n=1}^{\infty}b_n\,
ch^{NS}(h=n;\tau,z).
\label{A1} 
\end{eqnarray}
Here we have introduced the expansion \ba {1\over
2}\,\eta(\tau)\hskip-1mm\sum_{i=3,4}h_i(\tau)=-\sum_{n=1}b_nq^{n-1/8}
\ea 
and one can check by Maple that the expansion coefficients $b_n$ are positive integers for lower values of $n$. Actually one can prove that $b_n$ are positive integers for all values of $n$.\footnote{T.Eguchi and M.Jinzenji, 2008}

We also propose that elliptic genera of $A_{N-1}$ spaces are simply
$(N-1)$ times that of $A_1$ \ba Z_{{A_{N-1}}}(\tau, z)=(N-1){1\over
2}\left[\left(\theta_3(\tau,z)\over
\theta_3(\tau)\right)^2+\left(\theta_4(\tau,z)\over
\theta_4(\tau)\right)^2\right].
\label{Z A N-1}
\ea
Above construction (\ref{A1}) of $Z_{A_1}$ suggests that instead of
using the irreducible character $ch_0^{NS}(I=1/2)$ 
by itself we should use
its combination with (an infinity of) massive representations
defined by the R.H.S. of \eqn{A1}, 
which has a good modular property and is in fact invariant under
$\Gamma(2)$. We call this combination as the $\Gamma(2)$-invariant
completion of the massless representation 
and consider it as a conformal block in non-compact CFT.


\subsection{Theta-function identity}

It is a non-trivial problem to show 
that  for a given BPS representation 
of a superconformal algebra, it is possible 
to define its
$\Gamma(2)$-invariant completion
 uniquely by adding a suitable amount of
non-BPS representations.  According to our analysis this
seems possible when we impose suitable additional conditions:
all the massive contributions have 
their conformal dimensions above the gap, i.e. $h=n$ with
$n=1,2,\cdots$ and also occur with multiplicities of a definite sign.

The $\Gamma(2)$-invariant completion 
is the selection of  a topological part 
of massless representations; this may be easily
seen from the formula in the $\tilde{R}$ sector.
For instance we consider the decompositions
\ba
ch_0^{\tilde{R}}(I=0,\tau,z)&=&\left({\theta_3(\tau,z)\over
\theta_3(\tau)}\right)^2+h_3(\tau)\left({\theta_1(\tau,z)\over
\eta(\tau)}\right)^2,\label{masslessR1}\\
&=&\left({\theta_4(\tau,z)\over
\theta_4(\tau)}\right)^2+h_4(\tau)\left({\theta_1(\tau,z)\over
\eta(\tau)}\right)^2,\label{masslessR2} \ea 
We see at $z=0$, the 2nd
terms of (\ref{masslessR1}),(\ref{masslessR2}) vanish while the 1st
terms give the Witten index$\, =1$. Thus the 1st terms of (\ref{masslessR1}) and (\ref{masslessR2}) 
carry the topological information of the massless representation.
Our prescription is to identify the $\Gamma(2)$-invariant completion as  
 \ba
\left[ch^{\tilde{R}}_0(I=0;\tau,z)\right]_{inv}={1\over
2}\left[\left({\theta_3(\tau,z)\over
\theta_3(\tau)}\right)^2+\left({\theta_4(\tau,z)\over
\theta_4(\tau)}\right)^2\right]. 
\label{masslessR-top}
\ea 
Here we take the (GSO) projection 
$((\theta_3(\tau,z)/\theta_3(\tau))^2+\theta_4(\tau,z)/\theta_4(\tau))^2)/2$ since in
Ramond sector $q$-expansion is necesarily integer-powered. 
We do not adopt $(\theta_2(\tau,z)/\theta_2(\tau))^2$ 
since in this case associated 
massive representations start from $h=0$, 
i.e. below the threshold.

One of the most interesting examples of our analysis will be the case of the 
Appell function: 
It turns out that the desired completion is given by
\ba 
\hskip-3mm \left[{\cal K}_{2N}(\tau,z)\right]_{inv}
\equiv{1\over 4}{i\eta(\tau)^3\theta_1(\tau,2z)\over
\theta_1(\tau,z)^2}\left[\left({\theta_3(\tau,z)\over
\theta_3(\tau)}\right)^{2(N-1)}\hskip-3mm
+\left({\theta_4(\tau,z)\over
\theta_4(\tau)}\right)^{2(N-1)}\right]. \label{appell-top}\ea
Derivation will be given in Appendix B.


One can then plug the expression (\ref{appell-top}) into the orbifold formula
(\ref{orbifold}) and represent the elliptic genera for
$A_{N-1}$ spaces as 
\ba Z_{A_{N-1}}(\tau,z)&= &{1\over
4N}\sum_{a,b \in \bz_N}q^{{a^2\over 2}}e^{2\pi i
az}(-1)^{a+b}{\theta_1(\tau,{N-1\over N}z_{a,b}\,)
\theta_1(\tau,{2\over N}z_{a,b})\theta_1(\tau,z)
\over \theta_1(\tau,{1\over N}z_{a,b}\,)^3}\no\\
&& \hspace{5mm}\times  \left[\left({\theta_3(\tau,{1\over N}z_{a,b})\over
\theta_3(\tau)}\right)^{2(N-1)}\hskip-3mm
+\left({\theta_4(\tau,{1\over N}z_{a,b})\over
\theta_4(\tau)}\right)^{2(N-1)}\right]\,\label{improved-genus}
\ea where $z_{a,b}=z+a\tau+b$.

This appears to be a somewhat complicated formula. It turns out that rather strikingly this orbifold summation agrees
exactly with our proposed expression for $Z_{A_{N-1}}$
 \ba
\mbox{RHS of  }(\ref{improved-genus})={N-1\over 2}\left[
\left({\theta_3(\tau,z)\over
\theta_3(\tau)}\right)^2+\left({\theta_4(\tau,z)\over
\theta_4(\tau)}\right)^2\right].\label{identity} \ea 
We have proved
this identity for $N=2$ using the addition theorem of theta
functions and have checked its validity by Maple for lower values of  $N$. 
Mathematical proof for all values of $N$ has been given by Zagier \cite{Zagier}. 
Thus our approach seems altogether consistent: we have arrived at
the same expression (\ref{identity}) starting either from the decompactification of
K3 or the pairing of ${\cal N}=2$ minimal and Liouville theories. 
We have managed to construct holomorphic modular ($\Gamma(2)$)
invariant for a class of non-compact CY manifolds.

Actually the above identity (\ref{identity}) is
a special case of identities of theta functions 
\ba
&&{1\over 2N}\sum_{a,b \in \bz_N}q^{{a^2\over 2}}e^{2\pi i
az}(-1)^{a+b}{\theta_1(\tau,{N-1\over N}z_{a,b}\,) 
\theta_1(\tau,{2\over N}z_{a,b})\theta_1(\tau,z)
\over
\theta_1(\tau,{1\over N}z_{a,b}\,)^3}\,\left({\theta_i(\tau,{1\over N}z_{a,b})\over
\theta_i(\tau)}\right)^{2(N-1)}\hskip-8mm \no \\
&&\hskip3cm =(N-1)\left(\theta_i(\tau,z)\over \theta_i(\tau)\right)^2,
\hskip3cm i=2,3,4 \label{identity2}
\ea 
We note that the above  identities (\ref{identity2}) for $i=2,3,4$ transform into each other  under S and T transformations (more precisely under $SL(2;\bz)/\Gamma(2)=S_3$).

A mathematical proof of these identities (\ref{identity2}) has been found by D.Zagier \cite{Zagier}. We present his elegant proof using residue integrals in the next section.

\subsection{Proof of the theta-function identity}

Throughout this subsection we fix $\tau\in \bh $ (i.e.
$\mbox{Im}\,\tau >0$) and $N\ge2$. 
Also, for convenience we abbreviate
$\t(z)\equiv \t_1(\tau,z)/\t_1'(\tau,0)$ 
and $f_i(z)\equiv \t_i(\tau,z)/\t_i(\tau,0)$
($i=2,\,3,\,4$).  
We have $\t(z+a\tau+b)=(-1)^{a+b}q^{-a^2/2}y^{-a}\t(z)$
and similarly for $f_i(z)$, but with $(-1)^{a+b}$ being replaced by $(-1)^b$, 1 or
$(-1)^a$ for $i=2,\,3$ or 4, respectively. 
The identity \eqn{identity2} can therefore be rewritten as
\begin{eqnarray}
\frac1{2N}\,\sum_w\frac{\t((N-1)w)\;\t(2w)}{\t(Nw)\,\t(w)^3}\,f_i(w)^{2N-2}  
  &=& (N-1)\,\frac{f_i(z)^2}{\t(z)^2}\qquad(i=2,\,3,\,4), 
\label{theta-identity}
\end{eqnarray}
where the sum is over $w=(z+a\tau+b)/N$ 
with $a,\,b\in\Z_N$, or more invariantly
over $w\in E_\tau=\Bbb C/(\Z\tau+\Z)$ 
with $Nw=z$.

The proposed identity \eqn{theta-identity} is a special case of 
the more general ones:
\begin{eqnarray}
& & \frac1{2N}\;\sum_{Nw\,=\,z}\frac{\t((N-1)w)\;\t(2w)}{\t(Nw)}
 \,\t(w)^{2a-3}f_2(w)^{2b}f_3(w)^{2c}f_4(w)^{2d}  \nonumber \\ 
&& \hspace{2cm} = 
\left\{
\begin{array}{ll}
\displaystyle{ \frac{b\,f_2(z)^2+c\,f_3(z)^2+d\,f_4(z)^2}{\t(z)^2}}
& ~~\mbox{if} ~  a=0\\
1 & ~~ \mbox{if} ~ a=1 \\
0 & ~~ \mbox{if} ~ a\ge 2
\label{zagier-general}\end{array}
\right.
%
%
\label{theta-identity 2}
\end{eqnarray}
for any $a,\,b,\,c,\,d\ge0$ with $a+b+c+d=N-1$.

If we write $\wp(z)$ for $\wp(z;\tau)$
and observe that $f_i(z)^2/\t(z)^2=\wp(z)-e_i$ where $e_2\equiv
\wp(1/2)$, $e_3\equiv \wp((\tau+1)/2)$ 
and $e_4\equiv \wp(\tau/2)$, 
then we find that this identity \eqn{theta-identity 2}
follows from (and is in fact equivalent to) 
the following proposition: \\

\noindent
{\em \underline{Proposition}}

For $N\ge1$, let $F_N$ be the even elliptic function
$$ F_N(w)\=\frac{\t((N-1)w)\;\t(2w)\,\t(w)^{2N-5}}{\t(Nw)} , $$
and $P(X)=c_0X^{N-1}+c_1X^{N-2}+\text{\rm O}(X^{N-3})$ 
be a polynomial of degree $\le N-1$. Then
\begin{eqnarray}
 &&
 \frac1{2N}\;\sum_{Nw\,=\,z}
F_N(w)\,P(\wp(w))\=(N-1)\,c_0\,\wp(z)\,+\,c_1\,.
\label{theta proposition}
\end{eqnarray}


\noindent
{\bf [Proof]}

  Set $\z(z)\equiv \t'(z)/\t(z)$.  
This function satisfies $\z(z+a\tau+b)=\z(z)-2\pi ia$
for $a,\,b\in\Z$.  If we write the beginning of the Taylor expansion of $\t(z)$ at~0 as
$\t(z)=z+Az^3+\O(z^5)$ with $A=A(\tau)$ ($A$ is a multiple of $E_2(\tau)$), then we
have $\z(z)=z^{-1}+2Az+\O(z^3)$ and $\z'(z)=-z^{-2}+2A+\O(z^2)=-\wp(z)+2A$.
Fix $z\in\Bbb C$ (with $Nz\ne0$ in $E_\tau$) and define a function $t(w)$ by
$$t(w)\equiv \frac12\,\left\{\z(z+Nw)-\z(z-Nw) \right\}\,-\,\z((N-1)w)\,-\,\z(w)\,.$$ 
From the transformation law of $\z$ 
we find that $t(w)$ is elliptic.  
Therefore, by the residue theorem, we have
$$ \sum_{\a\in E_\tau} \,\text{Res}_{w=\a}\bigl(F_N(w)\,P(\wp(w))\,t(w)\,dw\bigr)\=0\,,$$
where the sum is over all singularities $\a\in\Bbb C/(\Z\tau+\Z)$ of $F_N(w)P(\wp(w))t(w)$.
These singularities occur only at $Nw=\pm z$ or $w=0$.  (The function $F_N(w)$ has further
simple poles at $\,Nw=0,\;w\ne0$, but $t(w)$ vanishes at these points, and the function
$t(w)$ has simple poles at $\,(N-1)w=0,\;w\ne0$, but $F_N(w)$ vanishes at these points.)
Since the residue of $t(w)$ at a point $w$ with $Nw=\pm z$ is $1/2N$ and $F_N(w)P(\wp(w))$ is
even, the identity above becomes
$$\frac1N\,\sum_{Nw\,=\,z}F_N(w)\,P(\wp(w)) \, + \, 
   \text{Res}_{w=0}\bigl(F_N(w)\,P(\wp(w))\,t(w)\,dw\bigr)\=0\,.$$
But for $w\to0$ we have
\begin{eqnarray}
F_N(w) &=& \frac{2(N-1)}N\,w^{2N-4}\,\bigl(1\,+\,A\,[(N-1)^2+2^2+2N-5-N^2]\,w^2
  +\O(w^4)\bigr) \nonumber \\ 
&=& \frac{N-1}N\,w^{2N-4}\,+\,\O(w^{2N})\,, 
\nonumber \\
  P(\wp(w)) &=& c_0\,\bigl(\frac1{w^2}+\O(w^2)\bigr)^{N-1}\,+\,
  c_1\,\bigl(\frac1{w^2}+\O(w^2)\bigr)^{N-2}\,
+\,\O\bigl(\frac1{w^2}\bigr)^{N-3}
  \nonumber \\
&=&
 \frac{c_0}{w^{2N-2}}\,+\,\frac{c_1}{w^{2N-4}}\,
+\,\O\bigl(\frac1{w^{2N-6}}\bigr)\,, \nonumber \\
  t(w) 
&=&
N\z'(z)\,w\,-\,\frac1{(N-1)w}\,-\,2A(N-1)w\,-\,
\frac1w\,-\,2Aw\,+\,\O(w^2) \nonumber
\\
  &= &-\,\frac N{N-1}\,w^{-1} \,-\,N\wp(z)\,w\,+\,\O(w^2)
\nonumber   ,
\end{eqnarray}
and hence $\;\text{Res}_{w=0}(F_N(w)\,P(\wp(w))\,t(w)\,dw)=-2(N-1)c_0\wp(z)-2c_1\,.\qquad\square$


\section{Summary}

When we consider a string theory on non-compact CY manifolds it is
described by a CFT possessing continuous as well as discrete
representations. Characters of representations of such CFT transform
in a peculiar manner under $S$ transformation as \ba
&& \mbox{discrete} \hskip6mm \rightarrow \hskip3mm \sum \mbox{discrete} +\int  \mbox{continuous}\no \\
&&\hskip2.1cm S\no \\
&&\mbox{continuous}  \hskip6mm \rightarrow  \hskip3mm \int \mbox{continuous}\no\\
&&\hskip2.6cm S\no \ea Mathematical nature of such transformation is
currently not well understood. We have found an empirical method of
constructing conformal blocks which have good modular behavior and
obtained elliptic genera of some non-compact CY manifolds. Our
method of construction of conformal blocks, however, is still
provisional and needs further studies.

From the geometrical point of view it is often difficult to define topological invariants for non-compact manifolds 
unambiguously and results tend to depend on the choice of boundary conditions. By our proposal  (\ref{A1 genera}), topological invariants of the $A_1$ space is predicted to be $\chi=1$, $\sigma=1$ and $\hat{A}=0$, respectively. These are more or less the standard values except that $A_1$ space is  topologically a cotangent bundle over $S^2$ and the Euler number may be considered as $\chi(S^2)$=2. 

 In our construction discrete representations describe homology classes of $H_2$ with compact support while the identity representation corresponds to the classes $H_0$, $H_4$. When we decouple gravity, we are left with one compact 2-cycle in the case of $A_1$ space and obtain $\chi$=1.  
We may take the point of view that our proposal is to impose good modular properties to fix the ambiguity of boundary conditions. In the case of complex 2-dimensions considered here the requirement of good modular behavior fixes the results uniquely.
We may, for  instance,  consider an alternative expression for the  topological part of ${\cal K}$ (\ref{appell-top}) where we replace 
$(\theta_3(\tau,z)/\theta_3(\tau))^{2(N-1)}+(\theta_4(\tau,z)/\theta_4(\tau))^{2(N-1)} $ by some symmetric polynomial in $\theta_3$ and $\theta_4$, such as, $(\theta_3(\tau,z)/\theta_3(\tau))^{2(N-2)}{(\theta_4(\tau,z)/\theta_4(\tau))}^2+(\theta_4(\tau,z)/\theta_4(\tau))^{2(N-2)}(\theta_3(\tau,z)/\theta_3(\tau))^2$. Such an ambiguity disappears after orbifold summation (\ref{improved-genus}) according to Zagier's formula (\ref{zagier-general}) and we obtain the same elliptic genera $Z_{A_{N-1}}$
 (\ref{identity}).

In the case of complex 4-dimensions, however,  
some ambiguity seems to remain. In complex 3-dimensions, on the other
hand, elliptic genera are known to be given by the product of the Euler
number times an universal function \cite{EOTY, KYY} 
and this continues to be the case in non-compact manifolds \cite{ES2}.

\section*{Acknowledgment}

T.E. would like to express his sincere gratitude to prof.A.Tsuchiya for his long friendship during the last 
decades and for sharing his enthusiasms and insights into  
mathematics and mathematical physics.

Research of T.E. and Y.S. 
is supported in part by Grant
in Aid from the Japan Ministry of Education,
Culture, Sports, Science and Technology. 
Research of A.T. was partially supported by the TMR programme 
HPRN-CT-2002-00225.


\section*{Appendix A: ~ $\cN=2$ Extended Characters}
\setcounter{equation}{0}
\def\theequation{A.\arabic{equation}}

We first list the  irreducible characters of ${\cal N}=2$ theory 
($q\equiv e^{2\pi i \tau}$, $y\equiv e^{2\pi i z}$): \\

\noindent {\bf (1) continuous representations}:\\
\ba ch^{NS}(h,Q;\tau,z)=q^{h-\frac{\hat{c}-1}{8}}y^Q{\theta_3(\tau,z)\over
\eta(\tau)^3}, ~~~ (h>{|Q|\over 2})\ea {\bf (2) discrete
representations}:\\
\ba ch_{dis}^{NS}(Q;\tau,z)=q^{{|Q|\over 2}-{\hat{c}-1\over 8}}y^Q{1\over
1+y^{sgn(Q)}q^{1/2}}{\theta_3(\tau,z)\over \eta(\tau)^3}\ea {\bf (3)
identity representation}:\\
\ba ch^{NS}_{id}(\tau,z)=q^{-{\hat{c}-1\over 8}}{1-q\over
(1+yq^{1/2})(1+y^{-1}q^{1/2})}{\theta_3(\tau,z)\over
\eta(\tau)^3}\ea Here $y=e^{2\pi iz}$ and $Q$ denotes the  $U(1)$
charge of ${\cal N}=2$ algebra.\\

 \noindent Extended characters are given by the sum over spectral flow of
 irreducible characters \eqn{irred-extended}:\\

\noindent {\bf (1) continuous representations}: \\
\ba  \chi^{NS}_{cont}(p,\alpha;\tau,z)
&=&q^{\frac{p^2}{2}}\Theta_{\alpha,N}(\tau,{2z\over
N}){\theta_3(\tau,z)\over \eta(\tau)^3}, ~~~
(h=
\frac{p^2}{2}+\frac{4\alpha^2+1}{4N})
\ea 
{\bf (2) discrete representations}: \ba
\chi^{NS}_{dis}(s,s+2r;\tau,z)
&=&\sum_{m\in \bz}{\left(yq^{N(m+{2r+1\over
2N})}\right)^{s-1\over N}\over 1+yq^{N(m+{2r+1\over
2N})}}y^{2(m+{2r+1\over 2N})}q^{N(m+{2r+1\over
2N})^2}{\theta_3(\tau,z)\over \eta(\tau)^3}
\ea 
{\bf (3) identity representations}:
 \ba \chi^{NS}_{id}(r;\tau,z)
&=& q^{-{1\over 4N}}\sum_{m\in \bz}q^{N(m+{r\over
N})^2+N(m+{2r-1\over 2N})}y^{2(m+{r\over N})+1}\no \\
&&\hskip0.7cm \times {1-q\over \left(1+yq^{N(m+{2r+1\over 2N})}\right)
\left(1+yq^{N(m+{2r-1\over 2N})}\right)}{\theta_3(\tau,z)\over
\eta(\tau)^3}
 \ea
Here $\Theta_{k,N}(\tau,z)$ is the theta function \ba
\Theta_{k,N}(\tau,z)=\sum_{m\in \bz} q^{N(m+{k\over
2N})^2}y^{N(m+{k\over 2N})}\ea Range of parameters $r,s$ are 
\ba
r\in \bz_N \,, \hskip3mm 1\le s\le N, \hskip3mm (s\in \bz). 
\ea

If we go to the Ramond sector with $(-1)^F$ insertion, one has \ba
\chi_{dis}^{\tilde{R}}(s,s+2r;\tau,z)=\sum_{m\in \bz}
{\left(yq^{N(m+{2r+1\over 2N})}\right)^{s-1\over N}\over
1-yq^{N(m+{2r+1\over 2N})}}y^{2(m+{2r+1\over 2N})}q^{N(m+{2r+1\over
2N})^2}{i\theta_1(\tau,z)\over \eta(\tau)^3}
\ea $r$ now takes half-integer values. We find discrete
representations $1\le s \le N$ with $r=-1/2$ carry a non-zero Witten
index \ba \chi_{dis}^{\tilde{R}}(s,s-1;\tau,z=0)=-1 \ea

Now we discuss S-transformation of extended characters. S-transform
of continuous representations remains essentially the same as the
Fourier transformation
\begin{eqnarray}
&& \chi_{cont}(p, m ; -\frac{1}{\tau})  
= \sqrt{\frac{2}{N}}\,
 \sum_{m' \in Z_{2N}}\, e^{-2\pi i \frac{mm'}{2N}}  \int_{0}^{\infty}dp'\,
 \cos(2\pi p p') \,
\chi_{cont}(p', m' ; \tau) .
\label{Liouville massive S}\nonumber \\
&&
\end{eqnarray}

S-transformation of discrete representations is given by 
\ba
&&\chi_{dis}^{NS}\left(s,m;-\frac{1}{\tau}\right)\no\\
&&\hskip-1cm =
\frac{1}{\sqrt{2N}}\sum_{m'\in
\bz_{2N}}\, e^{-2\pi i \frac{m m'}{2N}} \times \int_0^{\infty} dp'\, \frac{\cosh\left(2\pi
\frac{N-(s-1)}{\sqrt{2N}}p'\right) + 
e^{2\pi i\frac{m'}{2}} \cosh\left(2\pi
\frac{s-1}{\sqrt{2N}}p'\right)} {2\left| \cosh \pi
\left(\sqrt{\frac{N}{2}}p'+i\frac{m'}{2}\right) \right|^2} \,
\chi_{cont}^{NS}(p',m';\tau) \no\\ 
&& \hspace{-1cm} + \frac{i}{N}
\sum_{s'=1}^{N}\,\sum_{m'\in \bz_{2N}}\, e^{2\pi i
\frac{(s-1)(s'-1)-m m'}{2N}}\, \chi_{dis}^{NS}(s',m';\tau)- \frac{i}{2N} \sum_{m'\in \bz_{2N}}\,
e^{-2\pi i \frac{m m'}{2N}}\chi^{NS}_{cont}(p'=0,m';\tau),
 \label{S discrete} \ea where $m=s+2r$. 
The transformation of the identity
representation is 
\begin{eqnarray}
&&\hskip-1cm \chi_{id}(m;-\frac{1}{\tau})
=  \frac{1}{\sqrt{2N}} \,\sum_{m'\in Z_{2N}}\,
e^{-2\pi i \frac{mm'}{2N}}\int_{0}^{\infty} dp' \, 
\frac{\sinh\left(\pi {\cal Q} p'\right)\sinh(2\pi \frac{p'}{\cal Q})}
{|\cosh \, \pi (\frac{p'}{\cal Q}+i\frac{m'}{2})|^2}\, 
\chi_{cont}(p', m';\tau) 
\nonumber\\
&& \hskip1cm +\frac{2}{N}
\sum_{r'\in Z_{N}}\,\sum_{s'=2}^{N}\,
\sin(\frac{\pi (s'-1)}{N}) e^{-2\pi i \frac{m(s'+2r')}{2N}}
\, \chi_{dis}(s',r';\tau)
\label{Liouville identity S}
\end{eqnarray}

We refer the reader to \cite{ES1,ES2} for more complete 
discussions. 

 \section*{Appendix B: \hskip2mm $\Gamma$(2)-invariant Completion of Appell Function}
\setcounter{equation}{0}
\def\theequation{B.\arabic{equation}}

Let us consider the representation of ${\cal N}=4$ theory at general values of central charge $c=6k$ where $k$ is an arbitrary positive integer.
This theory  possesses an affine $SU(2)$ current of level $k$ which is given by a diagonal sum of 
level $k-1$ bosonic $SU(2)$ current and level 1 current made of fermion
bilinears. When we try to generalize
 the formula (\ref{masslessR1}), (\ref{masslessR2}) 
for a general level, we expect an expansion of the form
\ba
&&ch^{\tilde{R}}_0(I=0,\tau,z)=
\left({\theta_3(\tau,z)\over
\theta_3(\tau)}\right)^{2k}+\sum_{j=0}^{k-1}A_{3,j}(\tau)\chi^{(k-1)}_j(\tau,z){\theta_1(\tau,z)^2\over
\eta(\tau)^{3}}.
\label{expansion 3}
\ea
where $\chi^(k)_j$ denotes the $SU(2)_k$ 
character for spin $j/2$ representation
\begin{eqnarray}
 && 
\chi^{(k)}_{j}(\tau,z)= \frac{\Th{j+1}{k+2}(\tau,2z)-\Th{-j-1}{k+2}(\tau,2z)}{i\th_1(\tau,2z)}\equiv
2\,\frac{\Th{j+1}{k+2}^{(-)}(\tau,2z)}{i\th_1(\tau,2z)}~.
\label{SU(2) ch}
\end{eqnarray}
It turns out that expansion coefficients $A_{3,j}$  are given by
\ba
A_{3,j}(\tau)=2\widehat{H}^{(2(k+1))}_{j+1}(\tau)+a^{(k+1)}_{j+1}(\tau),\hskip3mm (j=0,1,\cdots,k-1)
\ea
where
\ba
&&\widehat{H}^{(k)}_s(\tau)\equiv {H^{(k)}_s(\tau)\over \Theta_{s,{k\over 2}}(\tau,\tau)},\hskip3mm H^{(k)}_s(\tau)\equiv \sum_{n\in \bz}{q^{{k\over 2}n(n+1)+(n+{1\over 2})s}\over 1-q^{k(n+{1\over 2})}}
\ea
and $a^{(k+1)}_{j+1}$ is expressed in terms of values of theta functions and $SU(2)$ characters at special points $z=r/2(k+1),\, (r=1,\cdots,2k+1)$. We shall prove the decomposition  formula  (\ref{expansion 3}) below.

 Similarly, one has the expansion
\ba
ch^{\tilde{R}}_0(I=0,\tau,z)=\left({\theta_4(\tau,z)\over \theta_4(\tau)}\right)^{2k}\hskip-2mm
+\sum_{j=0}^{k-1}A_{4,j}(\tau)\chi^{(k-1)}_j(\tau,z)
{\theta_1(\tau,z)^2\over \eta(\tau)^3}.
\label{expansion 4}
\ea
$A_{4,j}(\tau)$ is determined from $A_{3,j}(\tau)$ by taking the
T-transformation;
\begin{eqnarray}
 && A_{4,j}(\tau) = e^{2\pi i \frac{(j+1)^2}{4(k+1)}} A_{3,j}(\tau+1).
\end{eqnarray}

We keep the 1st terms of (\ref{expansion 3}), (\ref{expansion 4}), take the average and obtain
the $\Gamma(2)$-invariant completion
\ba
\left[ch^{\tilde{R}}_0(I=0,\tau,z)\right]_{inv}={1\over 2}
\left[\left({\theta_3(\tau,z)\over \theta_3(\tau)}\right)^{2k}
+\left({\theta_4(\tau,z)\over \theta_4(\tau)}\right)^{2k}\right].
\label{completion N=4}
\ea
If one recall the relation
\ba
(-1)^k q^{{k\over 4}}y^k ch_0^{NS}(I={k\over 2};\tau,z+{(\tau+1)\over 2})=
ch^{\tilde{R}}_0(k,I=0;\tau,z)
\ea
one finds
\ba
\left[ch^{NS}_0(I=\frac{k}{2},\tau,z)\right]_{inv}
={1 \over 2}
\left[
(-1)^k\left({\theta_1(\tau,z)\over \theta_3(\tau)}\right)^{2k}
+\left({\theta_2(\tau,z)\over \theta_4(\tau)}\right)^{2k}
\right].
\ea
Taking the half spectral flow 
$z\,\mapsto \, z-\frac{\tau}{2}-\frac{1}{2}$, 
we can rewrite $\frac{1}{2} \left(\eqn{expansion 3}
+ \eqn{expansion 4}\right)$ as 
\begin{eqnarray}
&& \hspace{-1cm}
{1 \over 2}
\left[(-1)^k \left({\theta_1(\tau,z)\over \theta_3(\tau)}\right)^{2k}
+\left({\theta_2(\tau,z)\over \theta_4(\tau)}\right)^{2k}
\right] =
ch^{NS}_0(I=k/2;\tau,z)+ \sum_{j=0}^{k-1}\sum_{n=0}^{\infty}
\, b_{j,n} \,  ch^{NS}(h^{(0)}_{j}+n ,j/2;\tau,z), \no \\
&&
\label{positivity}
\end{eqnarray}
where the coefficients $b_{j,n}$ are defined by the
$q$-expansion\footnote
    {$h^{(0)}_j$ in (\ref{NS massive term}) is found to take the
    following values;
$$
h^{(0)}_j= \left\{
\begin{array}{ll}
 \frac{k}{2}& ~~~ (j \equiv k ~ (\mbox{mod}\, 2)) \\
 \frac{k+1}{2}& ~~~ (j \equiv k-1  ~ (\mbox{mod}\, 2))
\end{array}
\right. ~.
$$
}
\begin{eqnarray}
 && q^{h^{(0)}_j - \frac{j(j+2)}{4(k+1)} -\frac{k^2}{4(k+1)}} \,
\sum_{n=0}^{\infty} b_{j,n} q^n 
= \frac{(-1)^{k+j+1}}{2} \left(A_{3,k-1-j}(\tau)
+ A_{4,k-1-j}(\tau)\right), 
\label{NS massive term}
\end{eqnarray}
and the ${\cal N}=4$ massive character of conformal weight $h$, spin
$j/2$ is given by
\begin{eqnarray}
ch^{NS}(h,j/2;\tau,z) \equiv q^{\frac{p^2}{2}}\chi^{(k-1)}_j(\tau,z)
\frac{\theta_3(\tau,z)^2}{\eta(\tau)^3}~, ~~~ 
h\equiv \frac{p^2}{2}+ \frac{j(j+2)}{4(k+1)}
+\frac{k^2}{4(k+1)}.
\label{massive character}
\end{eqnarray}
We again note that $b_{j,n}$ are non-negative 
integers as in \eqn{A1}. 
We have explicitly checked this for the cases $k=2,3,4$ by Maple.

Let us next study the relation of ${\cal N}=4$ character and Appell function.
Explicit form of ${\cal N}=4$ massless character in Ramond sector is given by\ba
&&ch_0^{\tilde{R}}(k,I=0;\tau,z)=-{i\theta_1(\tau,z)^2\over \eta(\tau)^3\theta_1(\tau,2z)}\,\sum_{n\in \bz}{1+yq^n\over 1-yq^n}q^{(k+1)n^2}y^{2(k+1)n}.
\ea
(This expression is obtained by slightly rewriting the original formula in  \cite{ET}. See (\ref{massless another exp}) in Appendix C).
Thus 
\ba
ch^{\tilde{R}}_0(k,I=0;\tau,z)&=&
- {i\theta_1(\tau,z)^2\over \eta(\tau)^3\theta_1(\tau,2z)}
\left({\cal K}_{2(k+1)}(\tau,z)-{\cal K}_{2(k+1)}(\tau,-z)\right)\no \\
&=&-2{i\theta_1(\tau,z)^2\over \eta(\tau)^3\theta_1(\tau,2z)}
\widehat{{\cal K}}_{2(k+1)}(\tau,z).
\label{N=4 and Appell}
\ea
By comparing \eqn{N=4 and Appell} with \eqn{completion N=4}
we obtain the invariant completion of Appell function
 \ba \hskip-1cm \left[\widehat{{\cal
K}}_{2N}(\tau,z)\right]_{inv}
\left(\equiv \left[{\cal K}_{2N}(\tau,z)\right]_{inv}\right)
\equiv{1\over 4}{i\eta(\tau)^3\theta_1(\tau,2z)\over
\theta_1(\tau,z)^2}\left[\left({\theta_3(\tau,z)\over
\theta_3(\tau)}\right)^{2(N-1)}\hskip-3mm
+\left({\theta_4(\tau,z)\over
\theta_4(\tau)}\right)^{2(N-1)}\right].
\ea

~
{\bf Proof of the relation (\ref{expansion 3})}\

Let us now prove the identity (\ref{expansion 3}).
Written in terms of the Appell function (\ref{expansion 3}) is expressed as
\begin{eqnarray}
\hcK_{2(k+1)}(\tau,z)={i\over 2}
{\eta(\tau)^3\theta_1(\tau,2z)\over \theta_1(\tau,z)^2} \left({\theta_3(\tau,z)
\over \theta_3(\tau)}\right)^{2k}+{i\over
2}\sum_{j=0}^{k-1}A_{3,j}(\tau)
\chi_{j}^{(k-1)}(\tau,z)\theta_1(\tau,2z)
\label{2nd decomposition}
\end{eqnarray}
As an intermediate step we first show the  following decomposition 
formula\\
\begin{eqnarray}
  \hcK_k(\tau,z) &=& G_k(\tau,z) + F_k(\tau,z)~.
\label{formula hcK}\\
G_k (\tau,z) &\equiv& \frac{i\eta(\tau)^3}{\th_1(\tau,z)}
\frac{\th_2(\tau,z)}{\th_2(\tau)} \prod_{\ell=0}^{k-1}
\frac{\th_4\left(\tau,-z+\frac{\ell}{k}\right)}
{\th_4 \left(\tau,\frac{\ell}{k}\right)}
\label{G},\\ 
 F_k(\tau,z)
&\equiv& \sum_{s=1}^{k-1}
\widehat{H}^{(k)}_s(\tau)
 \Th{s}{\frac{k}{2}}(\tau,2z).\label{F} 
\end{eqnarray}
(\ref{formula hcK}) is proved by checking that  $ \hcK_k(\tau,z) -F_k(\tau,z)$ has the same quasi periodicity and has the same zeros and poles in $z$ as the function $G_k(\tau,z)$. It is easy to see that both sides of (\ref{formula hcK}) have the same quasi-periodicity property.
We also note that the function $G_k(\tau,z)$ has poles at $z=n\tau+m$ while $F_k(\tau,z)$ is regular. These poles correspond to the zero of the denominators of the Appell function. We can check that $G_k$ and ${\widehat{\cal K}}_k$ have the same residues at these poles.  

Let us next show that $ \hcK_k(\tau,z) -F_k(\tau,z)$ vanishes at zeros of $G_k(\tau,z)$, i.e.  $z=z_{\ell}\equiv {\tau\over 2}+{\ell \over k}, \hskip1mm \ell=0,\cdots,k-1$. First  using the identity 
$$
\frac{1}{1-yq^n} = \sum_{s=0}^{k-1} \frac{(yq^n)^s}
{1-y^kq^{kn}}~,
$$
 we can rewrite $\hcK_k$ as  
 \begin{eqnarray}
 \hcK_k(\tau,z) 
&=& \frac{1}{2}\sum_{s=0}^{k-1} \sum_{n\in \bsz}
\left\{
\frac{\xi_{n,s}(\tau,z)}{1-y^k q^{kn}}
- \frac{\xi_{n,-s}(\tau,z)}{1-y^{-k} q^{-kn}}
\right\}
\label{evaluation hcK 1}
\end{eqnarray} 
 where 
\begin{eqnarray} 
  \xi_{n,s}(\tau,z) \equiv 
q^{-\frac{s^2}{2k}} \cdot 
q^{\frac{k}{2}\left(n+\frac{s}{k}\right)^2}
y^{k\left(n+\frac{s}{k}\right)}~.
\end{eqnarray}
 By setting $z=z_{\ell} \equiv \frac{\tau}{2}+ \frac{\ell}{k}$,
we obtain 
\begin{eqnarray}
 && \xi_{n,s}(\tau,z_{\ell})= e^{2\pi i \frac{\ell s}{k}}
 q^{\frac{k}{2}n(n+1) + s (n+\frac{1}{2})}~,
\end{eqnarray}
and thus, 
\begin{eqnarray}
 \hcK_k(\tau,z_{\ell}) &=& \frac{1}{2}
\sum_{s=0}^{k-1}\sum_{n\in \bsz} 
\left\{
\frac{q^{\frac{k}{2}n(n+1) + s (n+\frac{1}{2})}}
{1-q^{k(n+\frac{1}{2})}} e^{2\pi i \frac{\ell s}{k}}
- \frac{q^{\frac{k}{2}n(n+1) - s (n+\frac{1}{2})}}
{1-q^{-k(n+\frac{1}{2})}} e^{-2\pi i \frac{\ell s}{k}}
\right\} \nn
&=& \frac{1}{2}
\sum_{s=1}^{k-1}H^{(k)}_s(\tau)
\left(
e^{2\pi i \frac{\ell s}{k}} - e^{-2\pi i \frac{\ell s}{k}}
\right)~.
\end{eqnarray} 
 We also note 
 \begin{eqnarray}
 && \Th{s}{\frac{k}{2}}(\tau,2z_{\ell}) = 
q^{\frac{s^2}{2k}}\sum_{n\in \bsz} \xi_{n,s}(\tau,z_{\ell})
= e^{2\pi i \frac{\ell s}{k}} \Th{s}{\frac{k}{2}}(\tau,2z_0)
=  e^{2\pi i \frac{\ell s}{k}} \Th{s}{\frac{k}{2}}(\tau,\tau)~.
\end{eqnarray}
 Hence we obtain 
 \begin{eqnarray}
 && \hcK_k(\tau,z_{\ell}) = 
\sum_{s=1}^{k-1} H^{(k)}_s(\tau)
\frac{\Th{s}{\frac{k}{2}}(\tau,2z_{\ell})}
{\Th{s}{\frac{k}{2}}(\tau,\tau)} = F_k(\tau,z_{\ell}),
\end{eqnarray} 
where we used the identity $H^{(k)}_s(\tau)= - H^{(k)}_{k-s}(\tau)$.
Thus the difference $\hcK_k(\tau,z)-F_k(\tau,z)$ in fact vanishes at $z=z_{\ell}$.

Formula (\ref{2nd decomposition}) can then be derived by  
 modifying the function $G_k$ in (\ref{formula hcK}) so that it has a multiple zero at $z=1/2+\tau/2$ instead of simple zeros at $\{z_{\ell}\}$. 
Modification can be made by using a formula
 \begin{eqnarray}
 \hskip-3mm && \hspace{-1.5cm}
\left(\frac{\th_3(\tau,z)}{\th_3(\tau)}\right)^{2(K-1)} -
\prod_{j=1}^{K-1} 
\frac{\th_4\left(\tau,z+\frac{j}{2K}\right) 
\th_4\left(\tau, z-\frac{j}{2K}\right)} 
{\th_4\left(\tau,\frac{j}{2K}\right)^2} 
= \sum_{\ell=0}^{K-2} a^{(K)}_{\ell+1}(\tau)
\chi^{(K-2)}_{\ell}(\tau,z) \frac{\th_1(\tau,z)^2}
{\eta(\tau)^3}
\label{expansion G - cG 2}
\end{eqnarray} 
where $\chi_{\ell}^{(K-2)}(\tau,z)$ denotes the spin $\ell/2$ character of $SU(2)_{K-2}$.
Expansion coefficients $a_{\ell+1}^{(K)}(\tau)$ are determined below.
 
By combining (\ref{formula hcK}) and (\ref{expansion G - cG 2})  we arrive at 
(\ref{2nd decomposition}), (\ref{expansion 3}).\\

{\bf Determination of coefficients $a_{\ell}^{(K)}$}\\

Expansion coefficients $a_{\ell}^{(K)}(\tau)$ can be determined by comparing both sides of (\ref{expansion G - cG 2}) at $K-1$ points. By choosing the reference points $z={\tau\over 2}+{r\over 2K}\,(r=1,\cdots,K-1)$ which are the zeros of $G_{2K}$ we obtain a set of linear equations;
\begin{eqnarray}
 && \hspace{-1cm}
\sum_{s=1}^{K-1} a^{(K)}_s(\tau) \chi^{(K-2)}_{K-1-s}
\left(\tau,\frac{r}{2K}\right) = 
\frac{\th_2\left(\tau,\frac{r}{2K}\right)^{2(K-1)} \eta(\tau)^3}
{\th_3(\tau)^{2(K-1)} \th_4\left(\tau,\frac{r}{2K}\right)^2}
\equiv b_r(\tau) , ~ (r=1,\ldots,K-1).
\label{eq a_s}
\end{eqnarray}
Therefore, by means of the Cramer's formula we obtain
\begin{eqnarray}
 && a^{(K)}_j(\tau) = \frac{\mbox{det}\, B^{(j)}(\tau)}
{\mbox{det}\, B(\tau)}~, ~~~ (j=1,\ldots, K-1)
\label{sol a_s}
\end{eqnarray}
where $B(\tau)$, $B^{(j)}(\tau)$ are $(K-1)\times (K-1)$
matrices defined by
 \begin{eqnarray}
&& B(\tau) \equiv \left(B_{r,s}(\tau)\right)
_{1\leq r,s \leq K-1}~, ~~ B_{r,s}(\tau) \equiv
 \chi^{(K-2)}_{K-1-s}\left(\tau,\frac{r}{2K}\right)~, \nn
&& B^{(j)}(\tau) \equiv \left(
B_{r,1}(\tau), \cdots, B_{r,j-1}(\tau), b_r(\tau),
 B_{r,j+1}(\tau), \cdots, B_{r,K-1}(\tau)
\right)~.
\end{eqnarray}
We can simplify (\ref{sol a_s}) as follows; First we note 
\begin{eqnarray}
 && \chi^{(K-2)}_{s-1}(\tau, \frac{r}{K}) \equiv
\frac{2 \Th{s}{K}^{(-)}\left(\tau,\frac{r}{K}\right)}
{i \th_1\left(\tau,\frac{r}{K}\right)} = a_{r,s} 
\frac{\Th{s}{K}(\tau)}{g_r(\tau)}~,
\end{eqnarray}
where we set
\begin{eqnarray}
 && 
a_{r,s} \equiv \frac{\sin\left(\pi \frac{rs}{K}\right)}
{\sin \left(\pi \frac{r}{K}\right)}~, \label{ars} \\
&&
g_r(\tau) \equiv \frac{\th_1\left(\tau,\frac{r}{K}\right)}
 {2\sin\left(\pi\frac{r}{K}\right)}
\equiv q^{1/8} \prod_{m=1}^{\infty} (1-q^m)
 (1-e^{2\pi i \frac{r}{K}}q^m)(1-e^{-2\pi i \frac{r}{K}}q^m)~.
\label{gr}
\end{eqnarray}
Therefore,
\begin{eqnarray}
 && B_{r,s}(\tau) = a_{r,K-s}\frac{\Th{K-s}{K}(\tau)}
{g_r(\tau)} = (-1)^{r-s} a_{r,s}\frac{\Th{K-s}{K}(\tau)}
{g_r(\tau)}~, 
\end{eqnarray}
and the factor $\frac{\Th{K-s}{K}}{g_r}$ is factorized from
the determinant;
\begin{eqnarray}
 \det B(\tau) &=&  (-1)^{\frac{(K-1)(K-2)}{2}} 
\prod_{r,s=1}^{K-1}\frac{\Th{K-s}{K}(\tau)}{g_r(\tau)} \,
\det \hat{B}  \nn
&=& (-1)^{\frac{(K-1)(K-2)}{2}} 
\left(\frac{\eta(\tau)}{\eta(K\tau)}\right)^2
\prod_{s=1}^{K-1} 
\left(\frac{\Th{s}{K}(\tau)}{\eta(\tau)}\right) \,
\det \hat{B}~, 
\label{det B} \\
&& \mbox{where} ~ \hat{B} \equiv (a_{r,s}) ~. \nonumber
\end{eqnarray}
 In the second line we have used
$\prod_{r=1}^{K-1}g_r(\tau) = \eta(\tau)^{K-3} 
\eta(K\tau)^2$.
We similarly obtain
\begin{eqnarray}
 && \det B^{(j)} =
(-1)^{\frac{(K-1)(K-2)}{2}} 
\left(\frac{\eta(\tau)}{\eta(K\tau)}\right)^2
\prod_{s=1}^{K-1} 
\left(\frac{\Th{s}{K}(\tau)}{\eta(\tau)}\right) \,
\det \hat{B}^{(j)} ~, 
\end{eqnarray}
where we set
\begin{eqnarray}
\hskip-3mm && \hat{B}^{(j)}(\tau) \equiv \left(
a_{r,1}, \cdots, a_{r,j-1}, \hat{b}_{r,j}(\tau), 
a_{r,j+1}, \cdots
a_{r,K-1}
\right)~, \nn
\hskip-3mm && \hat{b}_{r,j}(\tau) \equiv b_r(\tau) \cdot 
\frac{(-1)^{r-1} g_r(\tau)}{\Th{K-j}{K}(\tau)} 
= \frac{(-1)^{r-1} \th_2\left(\tau,\frac{r}{2K}\right)^{2(K-1)}
\eta(\tau)^3}
{\th_3(\tau)^{2(K-1)} \th_4\left(\tau,\frac{r}{2K}\right)^2
\Th{K-j}{K}(\tau)
}\cdot \frac{\th_1\left(\tau,\frac{r}{K}\right)}
{2\sin\left(\pi \frac{r}{K}\right)}.
\label{hat b}
\end{eqnarray}
In this way we finally obtain a simplified formula;
\begin{eqnarray}
 && a^{(K)}_j(\tau)= \frac{\det \hat{B}^{(j)}(\tau)}
{\det \hat{B}}~.
\label{sol a_s 2}
 \end{eqnarray}

\section*{\bf Appendix C: \hskip2mm Modular property of ${\cal N}=4$ characters at higher level}
\setcounter{equation}{0}
\def\theequation{C.\arabic{equation}}

Let us recall the formula introduced by Miki \cite{Miki}  which is closely related to the Appell function, 
\ba
&&I(K,a,b;\tau,z)=\sum_{r\in \Z+1/2}e^{2\pi iar}\,{(yq^r)^{b}\over 1+yq^r}\,y^{Kr}\,
q^{Kr^2/2},
\ea
Its S-transform is given by
\ba
&&{i \over \tau}e^{-iK{\pi z^2\over \tau}}I(K,a,b;{-1\over \tau,}{z\over \tau})=\sum_{r\in \Z+a}e^{i\pi(r-a)}y^rq^{{r^2\over 2K}}{1\over \sqrt{K}}\int _{-\infty}^{+\infty}{e^{-2\pi b\left({p\over \sqrt{K}}+i{r\over K}\right)}\over 1+e^{-2\pi \left({p\over \sqrt{K}}+i{r\over K}\right)}}q^{{p^2\over 2}}\no \\
&&\hskip2cm +i\sum_{\stackrel{s\in \Z+1/2}{\delta(a,s)\not =0}}e^{i\pi (\delta(a,s)-a)}e^{2\pi i({K\over 2}-b)s}{(yq^s)^{\delta(a,s)}\over 1+yq^s}y^{Ks}q^{{K\over 2}s^2}\no \\
&&\hskip2cm +{i \over 2}\sum_{\stackrel{s\in \Z+1/2}{\delta(a,s)=0}}e^{-i\pi a}e^{2\pi i\left({K\over 2}-b\right)s}{1-yq^s\over 1+yq^s}y^{Ks}q^{{K\over 2}s^2}
\label{S-transform}\ea

Here $\delta(a,s)$ is a real number determined by the conditions
\ba
&&\delta(a,s)\equiv a-Ks \hskip2mm (\mbox{mod} \hskip2mm \Z),
\hskip3mm 0\le \delta(a,s)< 1.
\label{delta}\ea

\bigskip

We also recall  the higher level ${\cal N}=4$ massless character of
spin $\ell/2$ ($\ell=0,1,\ldots,k$)
\cite{ET} given by
\ba
&&\hskip-2cm ch^{NS}_0(k,\ell/2;\tau,z)=q^{-6k/24}q^{\ell/2}y^{\ell}\prod_{n=1}{(1+yq^{n-{1\over 2}})^2
(1+y^{-1}q^{n-{1\over 2}})^2
\over (1-q^n)^2(1-y^2q^n)(1-y^{-2}q^{n-1})}\no\\
&&\times\sum_mq^{(k+1)m^2+(\ell+1)m}y^{2(k+1)m}{1-(yq^{m+{1\over 2}})^{2(k-\ell+1)}\over
(1+yq^{m+{1\over 2}})^2}
\ea
which may be rewritten as
\ba
&&\hskip-2cm =-i{\theta_3(\tau, z)^2\over \eta(\tau)^3\theta_1(\tau, 2z)}
\sum_m q^{(k+1)m^2+(\ell+1)m+\ell+1/4-k/4}y^{2(k+1)m+\ell+1}{1-(yq^{m+{1\over 2}})^{2(k-\ell+1)}\over
(1+yq^{m+{1\over 2}})^2}\no \\
&&\hskip-2cm =-i{\theta_3(\tau,z)^2\over \eta(\tau)^3\theta_1(\tau,2z)}
\sum_m q^{(k+1)(m+1/2)^2}y^{2(k+1)(m+1/2)}(yq^{m+1/2})^{\ell-k}{1-(yq^{m+{1\over 2}})^{2(k-\ell+1)}\over
(1+yq^{m+{1\over 2}})^2}.\no \\
&&
\label{NSmassless}
\ea
Here the factor $(yq^{m+1/2})^{\ell-k}$ is extracted so that we can apply Miki's formula.
The last term in (\ref{NSmassless}) can be expanded as
\be
{1-(yq^{m+{1\over 2}})^{2(k-\ell+1)}\over
(1+yq^{m+{1\over 2}})^2}
=\sum_{i=0}^{2(k-\ell)+1}{(-1)^i(yq^{m+1/2})^i\over 1+yq^{m+1/2}}
\ee
Thus altogether the massless characters are written as
\ba
&&ch_0^{NS}(k,\ell;\tau,z)=-i{\theta_3(\tau,z)^2\over \eta(\tau)^3\theta_1(\tau,2z)}\no \\
&&\hskip-2cm \times\sum_m q^{(k+1)(m+1/2)^2}y^{2(k+1)(m+1/2)}
\sum_{i=0}^{2(k-\ell)+1}{(-1)^i(yq^{m+1/2})^{i+\ell-k}\over 1+yq^{m+1/2}}.
\label{massless another exp}
\ea
Terms in the 2nd line above have exactly the same form as the functions $I(2(k+1),a,b;\tau,z)$
and we have
\ba
&&ch_0^{NS}(k,\ell/2;\tau,z)=-i{\theta_3(\tau,z)^2\over \eta(\tau)^3\theta_1(\tau,2z)}\sum_{i=0}^{2(k-\ell)+1}I(K=2(k+1),i,i+\ell-k;\tau,z).\no \\
&&
\label{written as miki}\ea

Now we can apply the modular transformation law (\ref{S-transform}). We first note that the factor $\delta(a,s)$ 
of (\ref{delta}) vanishes
\be
\delta(a,s)=a-2(k+1)s\equiv 0 \hskip3mm \mbox{mod} \hskip2mm \Z
\ee
where $s=$1/2+integer.
Therefore only the 1st (continuous rep) and the 3rd term remain in (\ref{S-transform}).

We also note that the 3rd term has exactly the from as the  $\ell=k$ representation
\be
\mbox{3rd term}=(-1)^{a+b+k+1}\sum_m{1-yq^{m+1/2}\over 1+yq^{m+1/2}}y^{2(k+1)(m+1/2)}q^{(k+1)(m+1/2)^2}
\ee
Therefore we arrive at the result:
under the S-tranformation ${\cal N}=4$ massless characters produce only $\ell=k$ massless representation besides the continuous ones.
\be
ch_0^{NS}(k,\ell/2;-1/\tau,z/\tau)=\mbox{continuous reps} 
+(-1)^{\ell}(k-\ell+1)ch_0^{NS}(k,k/2,\tau,z).
\ee

Let us next examine the part of the continuous representations. 
We introduce the notation
\be
X_r\equiv e^{-2\pi\left({p\over \sqrt{2(k+1)}}+i{r\over 2(k+1)}\right)}
\ee
Continuous representations contained in (\ref{written as miki}) are given by
\ba
&&\sum_{i=0}^{2(k-\ell)+1}\sum_{r}(-1)^i(-1)^ry^rq^{r^2\over 4(k+1)}{1\over \sqrt{2(k+1)}}\int_{-\infty}^{\infty}{X_r^{i+\ell-k}\over 1+X_r}q^{p^2\over 2}.\no \\
\label{contell}\ea
After the sum over $i$
\ba
&&X_r^{\ell-k}\cdot \hskip-3mm \sum_{i=0}^{2(k-\ell)+1}(-1)^i{X_r^{i}\over 1+X_r}
={X_r^{\ell-k}-X_r^{k-\ell+2}\over (1+X_r)^2}
\label{contell3}\ea
 (\ref{contell}) is written as
\ba
\sum_r (-1)^r(y^r-y^{-r})q^{{r^2\over 4(k+1)}}{1\over \sqrt{2(k+1)}}\int {e^{-2\pi(\ell-k)\left({p\over \sqrt{2(k+1)}}+i{r\over 2(k+1)}\right)}\over \left[1+e^{-2\pi\left({p\over \sqrt{2(k+1)}}+i{r\over 2(k+1)}\right)}\right]^2}q^{{p^2\over 2}}dp.
\ea
Then by summing over $r$ modulo $2(k+1)$ the above formula is transformed to
\ba
&&\hskip-2cm \sum_{j=0}^{2k+1}{(-1)^j\over \sqrt{2(k+1)}}
\left[\Theta_{j,k+1}(\tau,2z)-\Theta_{-j,k+1}(\tau,2z)\right]\int {e^{-2\pi(\ell-k)\left({p\over \sqrt{2(k+1)}}+i{j\over 2(k+1)}\right)}\over \left[1+e^{-2\pi\left({p\over \sqrt{2(k+1)}}+i{j\over 2(k+1)}\right)}\right]^2}\,q^{{p^2\over 2}}dp \no \\
&&\hskip-1cm =\sum_{j=1}^{k}(-1)^j \sqrt{{2\over k+1}}
\left[\Theta_{j,k+1}(\tau,2z)-\Theta_{-j,k+1}(\tau,2z)\right] i\,Im\,\int {e^{-2\pi(\ell-k)\left({p\over \sqrt{2(k+1)}}+i{j\over 2(k+1)}\right)}\over \left[1+e^{-2\pi\left({p\over \sqrt{2(k+1)}}+i{j\over 2(k+1)}\right)}\right]^2}\,q^{{p^2\over 2}}dp .\label{contell2}\no\\
\ea
If one restores the overall factor $\theta_3(\tau,z)^2/\eta(\tau)^3\theta_1(\tau,z)$,
we recognize the combination $(\Theta_{j,k+1}(\tau,2z)-\Theta_{-j,k+1}(\tau,2z))/\theta_1(\tau,2z)$ as the level $k-1$ $SU(2)$ character. Thus (\ref{contell2}) has the form of  an integral over ${\cal N}=4$ continuous representations.

Then the modular transformation of massless representation is given by
\ba
&&\hskip-1cm ch_0^{NS}(k,\ell/2;-1/\tau,z/\tau)=e^{{2\pi i k z^2\over \tau}}\left[
\sum_{0\le \ell' \le k-1}\int_{-\infty}^{\infty} dp'
{\cal S}(\ell | p',\ell')ch^{NS}(k,p',\ell'/2;\tau,z)\right.\no \\
&&\left.\hskip3cm +(-1)^{\ell}(k-\ell+1)ch_0^{NS}(k,k/2;\tau,z)\right]
\ea
where the massive character is defind as in (\ref{massive character})
and the coefficient ${\cal S}(\ell |p',\ell')$ is given by
\begin{eqnarray}
{\cal S}(\ell |p',\ell')&=&(-1)^{\ell}\sqrt{{2\over (k+1)}} 
\,Im {e^{-2\pi(\ell'-k)\left({p'\over \sqrt{2(k+1)}}+i{\ell'+1\over 2(k+1)}\right)}\over \left[1+e^{-2\pi\left({p'\over \sqrt{2(k+1)}}+i{\ell+1\over 2(k+1)}\right)}\right]^2}
\no \\
 & \equiv & 
\frac{
e^{2\pi \frac{k-\ell +2}{\sqrt{2(k+1)}} p'} 
S^{(k-1)}_{\ell,\ell'}  
+ 2 e^{2\pi \frac{k-\ell +1}{\sqrt{2(k+1)}} p'} 
S^{(k-1)}_{\ell-1,\ell'}
+ e^{2\pi \frac{k-\ell}{\sqrt{2(k+1)}} p'} 
S^{(k-1)}_{\ell-2,\ell'}
}
{\left|
2 \cosh \pi \left(\frac{p'}{\sqrt{2(k+1)}}
+ i \frac{\ell'+1}{2(k+1)}
\right)\right|^4}.
\label{massless S coeff} 
\end{eqnarray}
Here 
$S^{(k-1)}_{\ell,\ell'} \equiv 
\sqrt{\frac{2}{k+1}} \sin \left(\pi 
\frac{(\ell+1)(\ell'+1)}{k+1}\right)$ is the modular coefficients of 
$SU(2)_{k-1}$.





\newpage
\begin{thebibliography}{100}

\bibitem{EST}
T.~Eguchi, Y.~Sugawara and A.~Taormina, 
"Liouville Field, Modular Forms and Elliptic Genera",
JHEP {\bf 0703}, 119 (2007)
[arXiv:hep-th/0611338]




\bibitem{FZZTZZ}
V.~Fateev, A.~B.~Zamolodchikov and A.~B.~Zamolodchikov,
arXiv:hep-th/0001012;
J.~Teschner,
arXiv:hep-th/0009138;
A.~B.~Zamolodchikov and A.~B.~Zamolodchikov,
arXiv:hep-th/0101152;


\bibitem{HK}
V.~Fateev, A.~B.~Zamolodchikov and A.~B.~Zamolodchikov,
unpublished;
A.~Giveon and D.~Kutasov,
JHEP {\bf 9910}, 034 (1999)
[arXiv:hep-th/9909110],
JHEP {\bf 0001}, 023 (2000)
[arXiv:hep-th/9911039];
K.~Hori and A.~Kapustin,
JHEP {\bf 0108}, 045 (2001)
[arXiv:hep-th/0104202].


\bibitem{ES1}
T.~Eguchi and Y.~Sugawara,
JHEP {\bf 0401}, 025 (2004)
[arXiv:hep-th/0311141],



\bibitem{ET}
T.~Eguchi and A.~Taormina,
Phys.\ Lett.\ B {\bf 200}, 315 (1988),
Phys.\ Lett.\ B {\bf 210}, 125 (1988).


\bibitem{Odake}
S.~Odake,
Mod.\ Phys.\ Lett.\ A {\bf 4}, 557 (1989);
Int.\ J.\ Mod.\ Phys.\ A {\bf 5}, 897 (1990).




\bibitem{Miki}
K.~Miki,
Int.\ J.\ Mod.\ Phys.\ A {\bf 5}, 1293 (1990).




\bibitem{IPT}
  D.~Israel, A.~Pakman and J.~Troost,
  JHEP {\bf 0404}, 043 (2004)
  [arXiv:hep-th/0402085].


\bibitem{Packman}
C.~Ahn, M.~Stanishkov and M.~Yamamoto,
Nucl.\ Phys.\ B {\bf 683}, 177 (2004)
[arXiv:hep-th/0311169],
JHEP {\bf 0407}, 057 (2004)
[arXiv:hep-th/0405274];
  D.~Israel, A.~Pakman and J.~Troost,
  Nucl.\ Phys.\ B {\bf 710}, 529 (2005)
  [arXiv:hep-th/0405259].
  A.~Fotopoulos, V.~Niarchos and N.~Prezas,
  Nucl.\ Phys.\ B {\bf 710}, 309 (2005)
  [arXiv:hep-th/0406017].
  K.~Hosomichi,
  arXiv:hep-th/0408172.


\bibitem{RS}
S.~Ribault and V.~Schomerus,
JHEP {\bf 0402}, 019 (2004)
[arXiv:hep-th/0310024].



\bibitem{OV}
H.~Ooguri and C.~Vafa,
Nucl.\ Phys.\ B {\bf 463}, 55 (1996)
[arXiv:hep-th/9511164].



\bibitem{GV}
  D.~Ghoshal and C.~Vafa,
  Nucl.\ Phys.\ B {\bf 453}, 121 (1995)
  [arXiv:hep-th/9506122].






\bibitem{Lerche}
W.~Lerche,
arXiv:hep-th/0006100.


\bibitem{HK2}
K.~Hori and A.~Kapustin,
JHEP {\bf 0211}, 038 (2002)
[arXiv:hep-th/0203147].


\bibitem{ES3}
  T.~Eguchi and Y.~Sugawara,
  JHEP {\bf 0501}, 027 (2005)
  [arXiv:hep-th/0411041].





\bibitem{Witten-CY/LG}
  E.~Witten,
  Nucl.\ Phys.\ B {\bf 403}, 159 (1993)
  [arXiv:hep-th/9301042].

\bibitem{Gepner}
  D.~Gepner,
  Nucl.\ Phys.\ B {\bf 296}, 757 (1988),
%
  Phys.\ Lett.\ B {\bf 199}, 380 (1987).


\bibitem{RY}
  F.~Ravanini and S.~K.~Yang,
  Phys.\ Lett.\ B {\bf 195}, 202 (1987).








\bibitem{Witten}
E.~Witten,
Int.\ J.\ Mod.\ Phys.\ A {\bf 9}, 4783 (1994)
[arXiv:hep-th/9304026].



\bibitem{ES2}
T.~Eguchi and Y.~Sugawara,
JHEP {\bf 0405}, 014 (2004)
[arXiv:hep-th/0403193].





\bibitem{STT}
A.~M.~Semikhatov, I.~Y.~Tipunin and A.~Taormina,
Commun. Math. Phys. {\bf 225}, 469 (2005)
[arXiv:math.qa/0311314].


\bibitem{Pol}
A. ~Polishchuk,
arXiv:math.AG/9810084.



\bibitem{KYY}
T.~Kawai, Y.~Yamada and S.~K.~Yang,
Nucl.\ Phys.\ B {\bf 414}, 191 (1994)
[arXiv:hep-th/9306096].


\bibitem{EOTY}
T.~Eguchi, H.~Ooguri, A.~Taormina and S.~K.~Yang,
Nucl.\ Phys.\ B {\bf 315}, 193 (1989).



\bibitem{Wendland}
K. Wendland, "Moduli Spaces of Unitary Conformal Field Theories", Ph.D. thesis, August 2000.


\bibitem{Seiberg}
  N.~Seiberg,
  Nucl.\ Phys.\ B {\bf 303}, 286 (1988).


\bibitem{Page}
  D.~N.~Page,
  Phys.\ Lett.\ B {\bf 80}, 55 (1978).

\bibitem{Zagier}
  D. Zagier, private communication, December 2006. 


\end{thebibliography}
\end{document}